\def\cm{{\rm\thinspace cm}}

\def\erg{{\rm\thinspace erg}}

\def\K{{\rm\thinspace K}}

\def\km{{\rm\thinspace km}}
\def\kpc{{\rm\thinspace kpc}}

\def\Mpc{{\rm\thinspace Mpc}}

\def\pc{{\rm\thinspace pc}}

\def\s{{\rm\thinspace s}}
\def\yr{{\rm\thinspace yr}}

\def\cmsqps{\hbox{$\cm^2\s^{-1}\,$}}

\def\ergpcmsqps{\hbox{$\erg\cm^{-2}\s^{-1}\,$}}

\def\kmps{\hbox{$\km\s^{-1}\,$}}

\def\kmpspMpc{\hbox{$\kmps\Mpc^{-1}$}}

\documentclass[usegraphicx]{mn2e}

\usepackage{amssymb}
\usepackage{mathptmx}

\include{defn}

\begin{document}

\title[The H$\alpha$ and X-ray emission in the Perseus cluster]
{The relationship between the optical H$\alpha$ filaments and the
X-ray emission in the core of the Perseus cluster}
\author[A.C. Fabian et al]{A.C. Fabian$^1$, J.S. Sanders$^1$, C.S.
Crawford$^1$, C.J. Conselice$^2$ J.S. Gallagher III$^3$ 
\newauthor and R.F.G. Wyse$^4$ \\
$^1$ Institute of Astronomy, Madingley Road, Cambridge CB3 0HA \\
$^2$ Department of Astronomy, Caltech, Pasadena CA, 91125, U.S.A. \\
$^3$ Department of Astronomy, University of Wisconsin-Madison, 475
North Charter Street, Madison, WI 53706-1582, U.S.A.\\
$^4$ Department of Physics \& Astronomy, Johns Hopkins University, 3400
North Charles Street, Baltimore, MD 21218, U.S.A.}

\maketitle

\begin{abstract}
NGC\,1275 in the centre of the Perseus cluster of galaxies, Abell\,426, is
surrounded by a spectacular filamentary H$\alpha$ nebula. Deep Chandra
X-ray imaging has revealed that the brighter outer filaments are also
detected in soft X-rays. This can be due to conduction and mixing of
the cold gas in the filaments with the hot, dense intracluster medium.
We show the correspondence of the filaments in both wavebands and draw
attention to the relationship of two prominent curved NW filaments to
an outer, buoyant radio bubble seen as a hole in the X-ray image.
There is a strong resemblance in the shape of the hole and the
disposition of the filaments to the behaviour of a large air bubble
rising in water. If this is a correct analogy, then the flow is
laminar and the intracluster gas around this radio source is not
turbulent. We obtain a limit on the viscosity of this gas.
\end{abstract}

\begin{figure}
\centering
\includegraphics[width=.95\columnwidth]{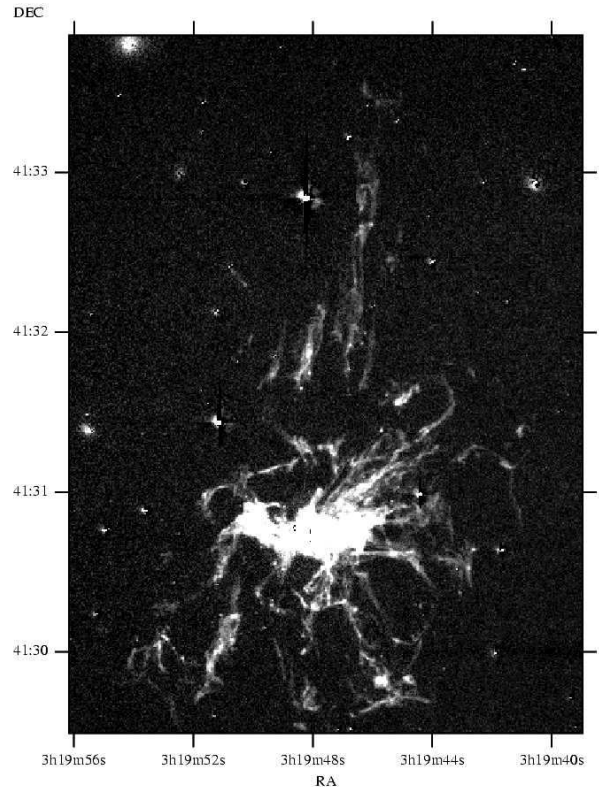}
\caption{WIYN telescope H$\alpha$ image of NGC\,1275 from
Conselice et al (2001). }
\label{fig:2t}
\end{figure}

\section{Introduction}

The giant galaxy NGC\,1275 at the centre of the Perseus cluster has
long been known to be surrounded by a large H$\alpha$ nebulosity
(Minkowski 1957; Lynds 1970). The blobs and filaments of the
nebulosity are embedded in dense hot intracluster medium and have low
ionization. Neither the origin of the filaments or the source of
ionization is clear, despite many studies (Kent \&
Sargent 1979; Cowie et al 1983; Johnstone et al 1988; Heckman et al
1989; McNamara, O'Connell \& Sarazin 1996; Sabra et al
2002). They may have formed by cooling of the intracluster medium
(Fabian 1994).

Recent deep H$\alpha$ obtained with the WIYN telescope \footnote{The
WIYN Observatory is a joint facility of the University of
Wisconsin-Madison, Indiana University, Yale University, and the
National Optical Astronomy Observatory.} (Conselice et al 2001) and
X-ray images (Fabian et al 2000; 2003) on arcsec resolution have now
been carried out. Here we compare features in both wavebands to see
what is common to both and whether it helps understand the origin of
the nebulosity. Some corresponding features around the radio source
have been seen in A2597 (Koekemoer et al 1999), A2052 (Blanton et al
2003) and as an 80~kpc long trail in A1795 (Fabian et al 2000).

\begin{figure*}
\centering
\includegraphics[width=1\columnwidth]{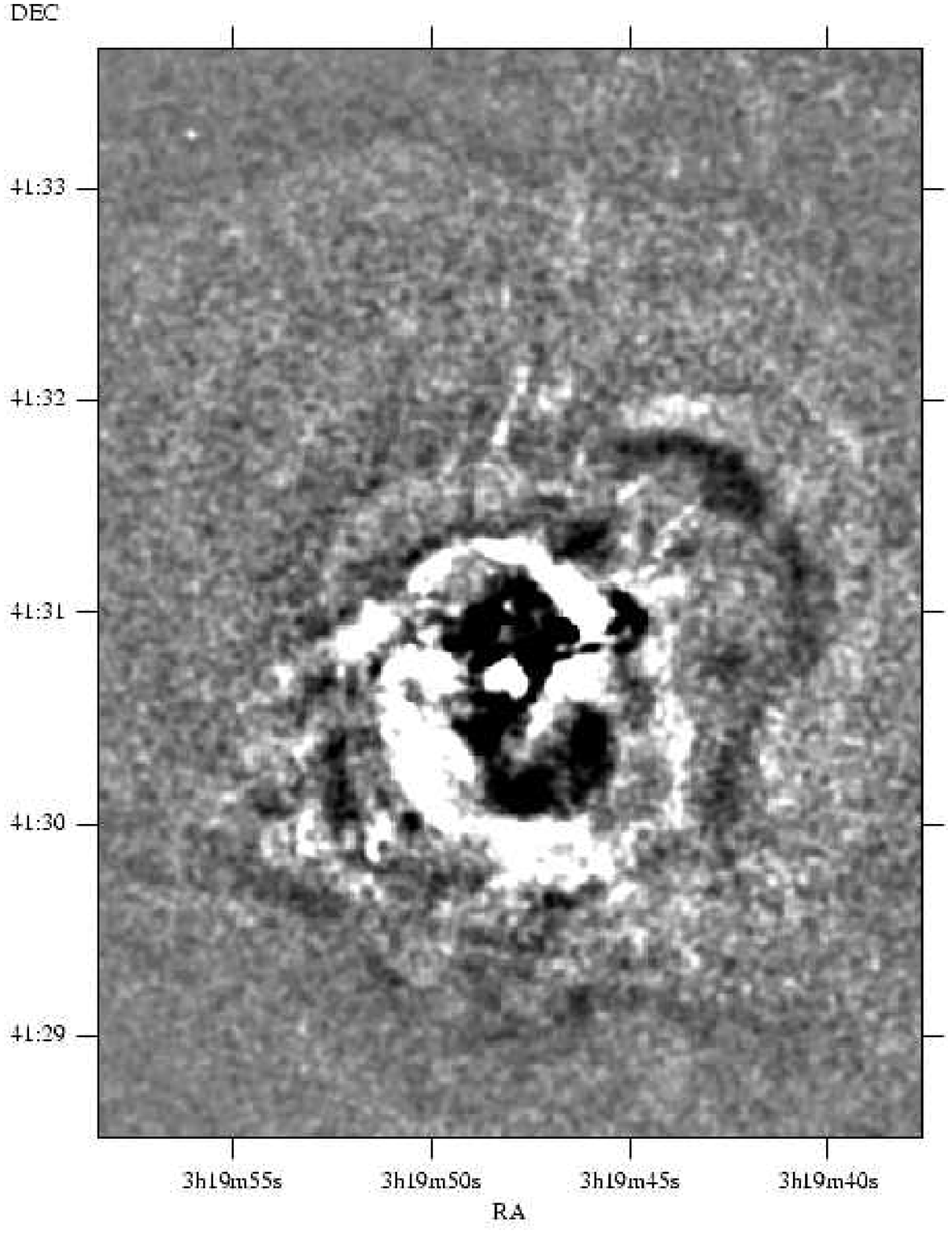}
\includegraphics[width=1\columnwidth]{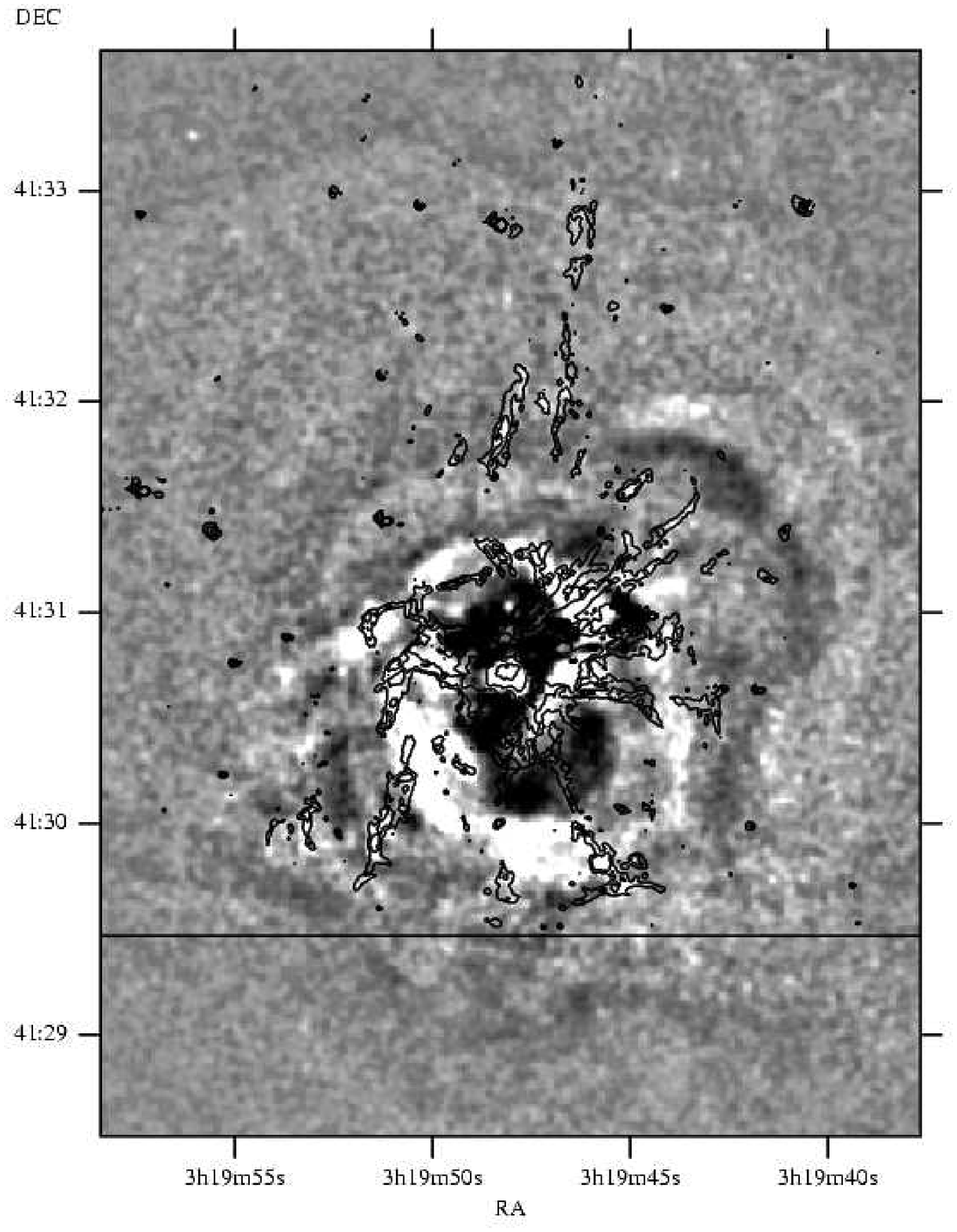}
\caption{Left: Deep unsharp mask, 0.3--1.5~keV (0.98 arcsec gaussian smoothed
with 9.8 arcsec gaussian smoothed image subtracted) X-ray image.
Regions brighter than the local average are light, those fainter are
dark.
Right: overlay of contours from the H$\alpha$ image on the unsharp mask
X-ray image. The horizontal line indicates the edge of the H$\alpha$
image.}
\label{fig:2t}
\end{figure*}

Large H$\alpha$ and Ly$\alpha$ nebulosities (the NGC\,1275 nebulosity
is also seen in Ly$\alpha$; Fabian et al 1984) are often found around
powerful radio sources over a wide range of redshifts (McCarthy 1993;
Reuland et al 2003) and are often considered to be the aftermath of
merger events, although the evidence is not conclusive. Fabian et al
(1985) noted that they are seen in many central galaxies in cooling
flow clusters and do not there correlate in luminosity with present
radio luminosity (see e.g. Crawford et al 1995). NGC\,1275 has both a
moderately powerful radio source (Pedlar et al 1990) and a strong
cooling flow.

The redshift of NGC\,1275 is 0.0183, which means that 1~kpc
corresponds to about 2.7 arcsec (using $H_0=70\kmpspMpc$).

\section{The H$\alpha$ and X-ray images}

The H$\alpha$ image is shown in Fig.~1 (Conselice et al 2001). The
total extent is about 4 arcmin or 80~kpc. There is no significant
filamentary emission beyond that shown in the rest of that image or in
one taken to the south. 

Filamentary structures are clearly seen in the soft X-ray image
(Fabian et al 2003) but are difficult to present simply due to the
large dynamic range of the underlying X-ray emission. We have
therefore used an unsharp-mask technique to bring them out (Fig.~2a).
The image is accompanied by one on which contours of the H$\alpha$
nebulosity have been overlaid (Fig.~2b). This enables filamentary
features to be compared by inspection. The main X-ray structures in
Fig.~2 (left) are the bright rims around the inner radio source, which
appear as the N and S radio holes (dark) coinciding with the radio
lobes (Fabian et al 2000, 2003).

Obvious correspondence between H$\alpha$ and soft X-ray features is
seen to the north, where 3 major strands of H$\alpha$ emission are
seen beginning about 1 arcmin to the N (RA 3 19 48.5, Dec 41 31 45)
of the nucleus. A 0.5 arcmin N-S stretch of the middle strand is most
obvious. A very faint structure is tentatively seen extending further
to the N in the X-ray image, bending to the W then S at the top.

At p.a. 140 deg and also beyond about 1 arcmin from the nucleus, other
H$\alpha$ filaments (3 19 51.4, 41 29 56) overlap patches of bright
soft X-ray emission. This is the `blue loop' of Sandage (1972) where
young stars may be forming away from the body of the galaxy (Conselice
et al 2001). At about 200 deg a radial filament extends from over
0.5--1.2~arcmin with some X-ray emission seen across the southern
radio hole and also at its outer, tangentially-oriented, end (3 19
45.4, 41, 29 45).

\begin{figure*}
\centering
\includegraphics[width=0.8\columnwidth]{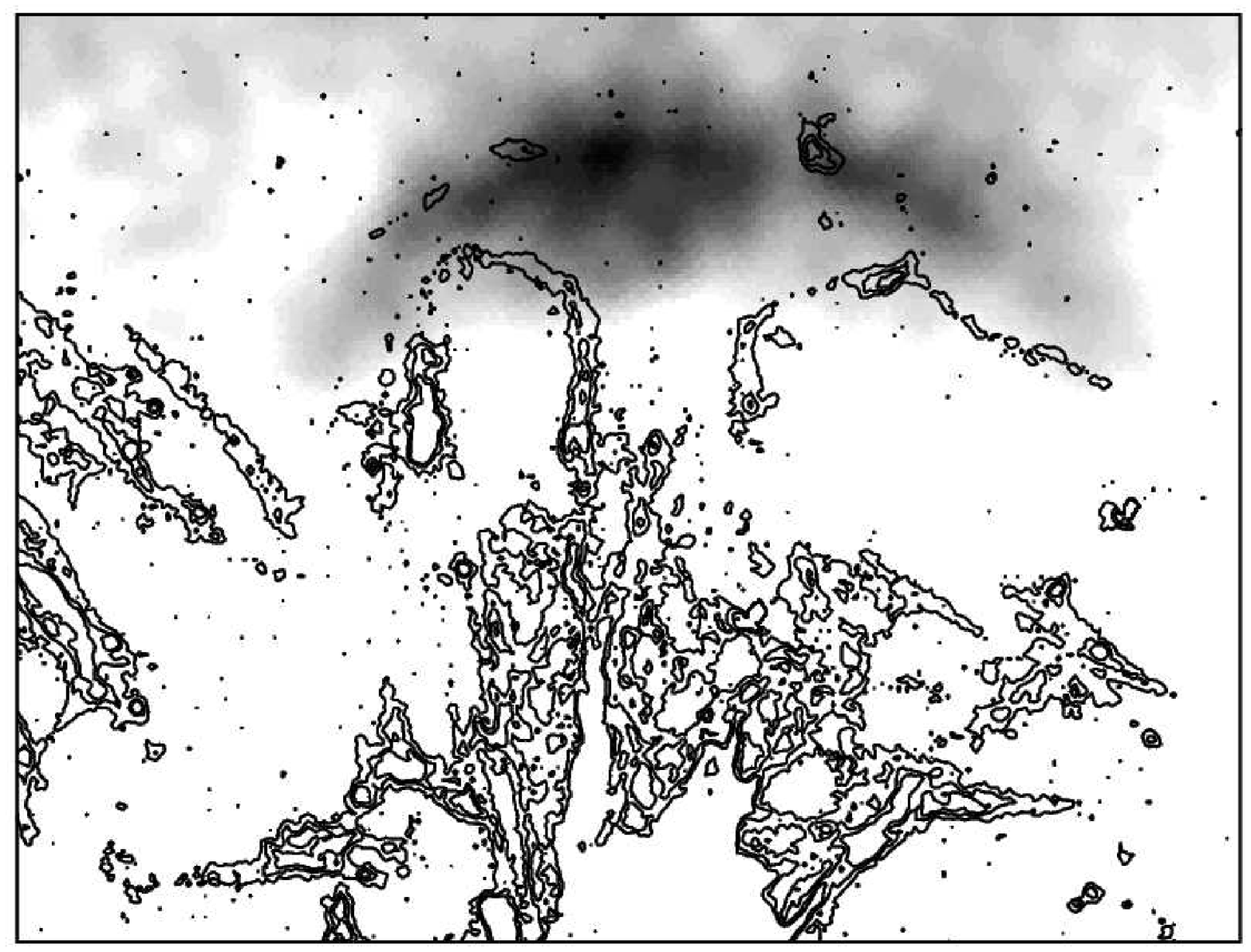}\hfil
\includegraphics[width=0.8\columnwidth]{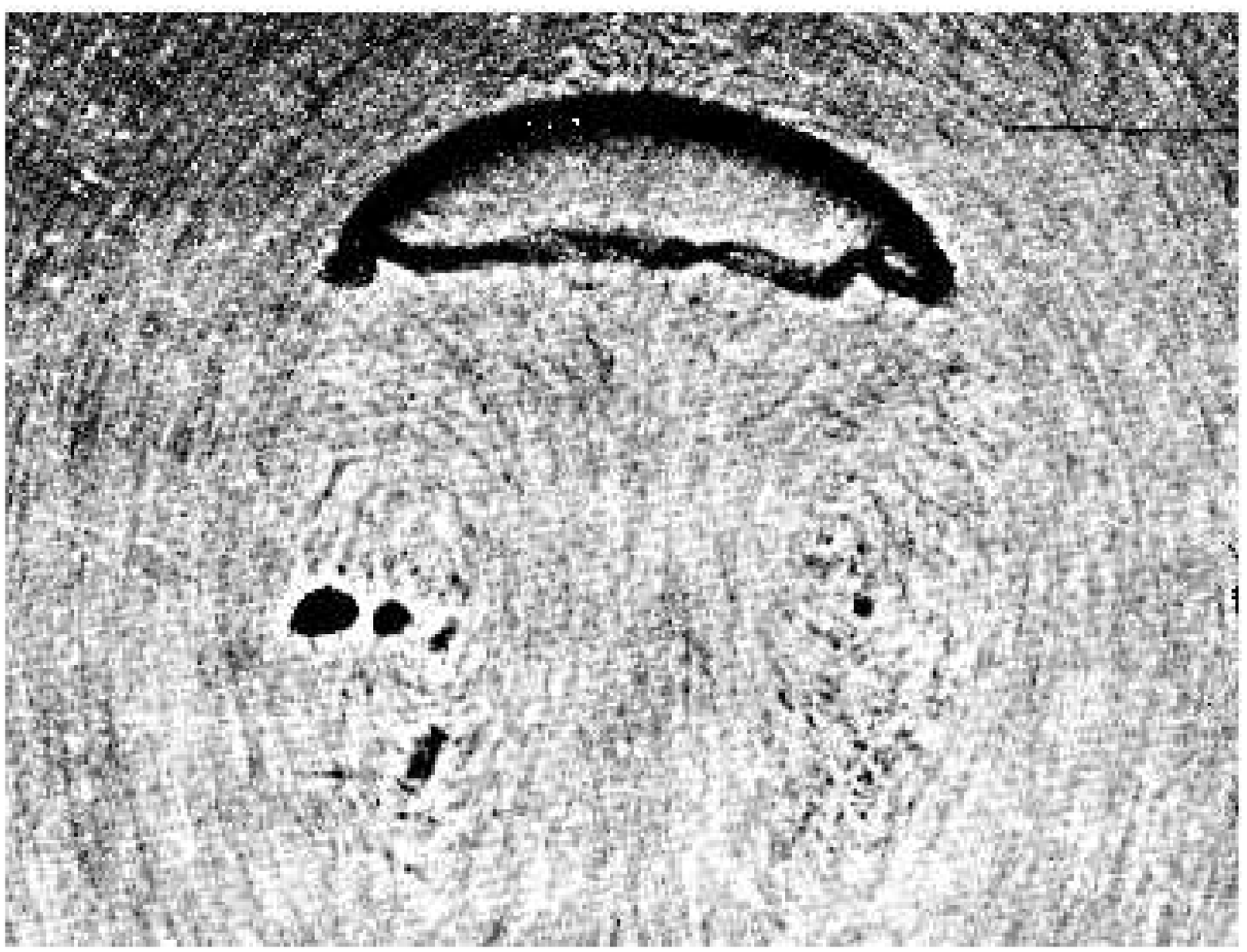}
\caption{Left; a) Contours from the H$\alpha$ image rotated 50 deg
counter-clockwise overlaid on a lightly smoothed X-ray image with
levels chosen to emphasise the outer rising bubble. Right; b) A rising
air bubble in water from Bachelor (1967, Plate 15c). This is
effectively a two-dimensional bubble since it is confined between
closely spaced parallel plates, but shows the cross-section and flow
pattern well (see Van Dyke 1982 for a full bubble). The effect of
surface tension is negligible in b) (Bachelor 1967). Magnetic fields
inside the bubble in a) are subequipartition (Fabian et al 2002), so
unlikely to influence strongly the comparison here. A gas bubble in
water may oscillate from side to side while rising. }
\label{fig:2t}
\end{figure*}

At p.a. 300 deg lies an H$\alpha$ filament which bends back on itself
(3 19 43.7, 41 31 27; called the `horseshoe' by Conselice et al 2001).
This feature is seen well in the soft X-ray image. Its projected
location lies just inside the outer, buoyant radio bubble and is
discussed further in the next section.

A careful inspection of the images reveals that most of the brighter
H$\alpha$ filaments, with intensity greater than about
$2\times10^{-15}\ergpcmsqps$~arcsec$^{-2}$ (Table 1 in Conselice et al
2001), at radii of 30 arcsec or more, show very similar structures in
soft X-rays (e.g. the horseshoe, Fig.~2. It is therefore possible that
all the outer H$\alpha$ filaments will have a soft X-ray counterpart
in even deeper X-ray images. The X-ray surface brightness of the
detected filaments is at the limit of sensitivity for fine structures
in the deep Chandra image. It is also in agreement with that expected
if a few 100~pc of $4\times 10^7\K$ gas along the line of sight is
replaced by gas at $1-2\times 10^7\K$ gas at the same pressure. (We
have fitted a thermal plasma model to the X-ray spectrum of the
northern filament and obtain a gas temperature of about $10^7\K$.)

At smaller radii the X-ray emission is strongly affected by the
`holes' coinciding with the radio lobes. A fine straight optical
filament about 30 arcsec S of the nucleus is seen in X-rays but little
else corresponds well. Note that the gas associated with the
high-velocity system in the foreground of the Per A nucleus
(redshifted by $3000\kmps$; see discussion by Conselice et al 2001)
causes soft X-ray absorption in the region extending to about
20~arcsec to the NE of the nucleus (Fabian et al 2000).

Projection effects mean that we cannot be precise about the physical
arrangement of the filaments. It is possible that most avoid the radio
lobes themselves, and the outer filaments appear to avoid the broad,
probably weakly-shocked, thick X-ray rim to the lobes. This thick rim,
with outer radius 1~arcmin, extends beyond the thinner, bright, soft
X-ray rim. This is different from the behaviour of the filaments in
A2597 (Koekemoer et al 1999) and A2052 (Blanton et al 2003) where the
filaments appear to surround the radio lobes more closely.

To assess the intensity of the soft X-ray emission from the filaments
we have examined the prominent filaments about one arcmin N of the
nucleus. The X-ray surface brightness there is on average about
$16.8$~ct~arcsec$^{-2},$ which for a plasma at $10^7\K$ corresponds to
an intensity of $9\times 10^{-16}\ergpcmsqps$~arcsec$^{-2}.$ This is
region 18 in Conselice et al (2001), where they find an H$\alpha$
intensity of $5\times 10^{-15}\ergpcmsqps$~arcsec$^{-2}$. Since
recombination implies that the Ly$\alpha$ intensity is about 20 times
larger still (see e.g. Donahue \& Voit 1991), the filaments
predominantly radiate in the UV/optical, with the X-ray emission being
an energetically minor component.

\section{The horseshoe}

We now consider the horseshoe-shaped filament at p.a. 300 deg. The
interesting aspect here is that it turns back just before the outer
buoyant bubble (Fig.~3a). The situation strongly resembles that seen
when a large air bubble rises in water (Fig.~3b; Bachelor 1967). The
bubble becomes a spherical cap trailing a toroidal flow pattern. The
Perseus bubble is cap-shaped and the H$\alpha$ filaments could be
revealing the flow pattern. In particular the filaments bend in
opposite directions either side of the centre of the bubble, as
expected. They represent natural streamlines tracing the flow in the
gas.  If the filaments are not in the plane of the Sky then doppler
shifts could provide spectroscopic confirmation of the flow pattern.

From the intensity of the H$\alpha$ emission, we estimate that the
emission region, if at $10^4\K$ and in pressure equilibrium with the
hot gas, is less than 0.1~pc thick. This means that the volume filling
fraction is very small; if part of a filament is  represented by a
cube of side one arcsec, or 300~pc, then the filling fraction is less
than $3\times 10^{-4}$. Consequently, and despite the large density
contrast, the filaments will be blown about by the hot gas and serve
as good tracers of its flow on kpc scales or more.  

This interpretation means that the rising bubble is drawing up
H$\alpha$ gas behind it. The projected temperature structure of the
hot intracluster gas is fairly smooth in the region so providing no
further information. The dragging up of colder, inner gas has
previously been discussed within the context of the inner hot
intracluster gas around the radio source in the Virgo cluster around
M87 (B\"ohringer et al 1995; Churazov et al 2001). In the Perseus cluster we
appear to have a dramatic demonstration of the flow pattern.

The flow appears to be laminar, indicating that the effective Reynolds
number, ${\cal R}=uL/\nu$, where $u, L$ and $\nu$ are the velocity,
cap radius and viscosity respectively, is less than about 1000.
(See Van Dyke 1982 for a photograph of a turbulent wake, from Wegener
\& Parlange 1973 where spherical-cap bubbles are reviewed.) This
yields a lower limit on the viscosity of the intracluster medium of
$\nu=uL/1000$ or $\nu>4\times 10^{27}\cmsqps$ assuming $700\kmps$ for
$u$ (Fabian et al 2002) and 20~kpc for $L$. The ion viscosity from the
formula of Spitzer (1962) and Braginskii (1958)
$\nu=10^8T^{5/2}n^{-1}=6\times 10^{28}\cmsqps$. 

The inclusion of magnetic fields makes the viscosity anisotropic. If
the net effect is to reduce it then the above argument means that the
factor cannot be more than ten. Faraday rotation measure maps of
cluster cores indicate magnetic field coherence lengths of 5--10~kpc
(Carilli \& Taylor 2002). Faraday rotation is so strong in the Perseus
core that the radio source is completely depolarized (Pedlar et al
1990). If this is due to strong, and dynamically dominant, magnetic
fields in the innermost intracluster medium, then these will surely
influence the effective viscosity and the flow.

Numerical simulations of the intracluster medium will need to ensure
that the gas has an effective viscosity at this level (the simulations
of e.g. Reynolds et al 2002 show turbulence) and to confirm that the
flow is indeed laminar. Viscosity can have an important effect in
dissipating the sound energy produced by the formation of the bubbles
(Fabian et al 2003; see references therein for a discussion of the
heating of the intracluster medium by a central active nucleus).

\section{Discussion}

The brighter H$\alpha$ filaments are detected in soft X-rays. 
The X-ray emission could be due to thermal conduction making the
intracluster gas adjacent to the filaments cooler, i.e. to conductive
condensation onto the filament, or to conductive evaporation of the
filament. This problem has been studied by B\"ohringer \& Fabian
(1989) for the case of a spherical cloud embedded in a medium similar
to that of the Perseus cluster (see also Sparks, Macchetto \& Golombek
1989).  

The results there (Table 2; B\"ohringer \& Fabian 1989) indicate that
the temperature drops around a cloud out to $\sim10f^{1/2}\kpc$,
where $f$ is a factor, assumed constant, by which conduction is
suppressed below the Spitzer (1957) value. Since they are observed to
be only 1--2 arcsec across, $10f^{1/2} < 0.5$ and $f<3\times 10^{-3}$.
An isolated H$\alpha$-emitting cloud can only be 10~pc or less in
radius. This would however be undetectable and a filament is
presumably a low filling-factor chain of small clouds. We assume that
the magnetic-field structure of the clouds is responsible for the required
suppression of conduction and helps give a cloud its coherence.

The timescale for attaining a complete equilibrium conductive boundary
layer is measured in Gyr (B\"ohringer \& Fabian 1989), which is much longer
than the likely age of the observed filaments (a few $10^7\yr$ from
the flow time). The observed clouds will therefore be in the early
evaporative phase, slowly losing gas to the surrounding intracluster
medium. The mismatch in timescales is so large that we do not expect
that the precise geometry of the clouds will affect this conclusion.
If there is a large relative velocity between the filaments and the
surrounding gas then forced mixing may be taking place. The immediate
$\sim100\pc$ surroundings of a filament is X-ray bright because the
gas is cooled (and thus denser) there by conduction or mixing with the
cold gas.

Being only a minor component energetically, the soft X-rays cannot be
the source of ionization of the filaments. (The short cooling time of
the H$\alpha$ emitting gas means that they must be continuously
ionized.) A conductive interface does not help either (B\"ohringer \&
Fabian 1989). The formation of massive stars within filaments (such as
are seen in the region of the blue loop) would potentially provide
sources of photoionization, and is a possibility which can be
observationally tested.

The northern H$\alpha$ filaments stop at the weak shock front
discovered in the deep Chandra image (Fabian et al 2003). Also the
H$\alpha$ emission seems weaker in Fig.~1 around the inner bubble
rims (e.g. to the SE). This could be the result of strong heating or
disruption in the shocked gas. It further supports the weak shock
interpretation of Fabian et al (2003) for the outer front seen in the
X-ray image.

The long, outer H$\alpha$ filaments to the N could result from being
drawn out by an older outer bubble, which is no longer detectable as a
hole in X-rays. This explains why it is so radial. Its age, from the
flow time, would be at least $7\times 10^7\yr$. If correct, it would
increase the viscosity limit in Section 3 by a factor of a few. It
would also remove any concern that the flow has not yet reached 
equilibrium (the outer bubble is about one eddy length from the
centre). The straightness of this and other filaments further
testifies against the hot medium being turbulent. The level of
turbulent, or other, motions in the hot gas will be measured by the
bolometer on ASTRO-E2, when it is launched. No obvious filaments go to
the south, where there is another, less well-defined, outer bubble
(Fabian et al 2001, 2003). The process presumably depends on the
quantity of cold gas available for being drawn out.

A strong possibility emerges therefore that all the filaments are due to
buoyant radio bubbles drawing out cold gas from a central reservoir.
The outflow of the radio source is responsible (Conselice et al 2001).
The filaments cover such a large range of cooling time within the hot
gas (to more than 1~Gyr) that it is unlikely that they originate {\it
in situ} by cooling from that gas. Where the cold gas reservoir comes
from is uncertain. It may accumulate from residual cooling of the
intracluster medium at the centre or from mergers. A detailed
spectroscopic comparison of abundances in the filaments and
intracluster gas (as suggested by Conselice et al 2001) could be
decisive.

The action of radio bubbles drawing out cold dusty filaments to large
radii along the radio axis could contribute to the alignment effect
seen in distant radio galaxies (McCarthy 1993).


\section*{Acknowledgements}
We thank the referee, Sebastian Heinz, for helpful comments. ACF and
CSC thank the Royal Society for support. CJC acknowledges support from
an NSF Astronomy and Astrophysics Fellowship. JSG, and RFGW thank the
US National Science Foundation for partial support of this work
through grants AST98-03018, and AST98-04706.

\end{document}